\documentclass[%
 reprint,
 superscriptaddress,
 amsmath,amssymb,
 aps,
 pra,
]{revtex4-2}

\usepackage{graphicx}
\usepackage{dcolumn}
\usepackage{bm}
\usepackage{subcaption}
\usepackage{orcidlink}
\usepackage[justification=raggedright,singlelinecheck=false]{caption}
\usepackage[percent]{overpic}
\hypersetup{hidelinks}
\begin{document}

\title{Field-driven attosecond deflection of electron beams at the position of planar foils}
\date{\today}

\author{Xiaofan Gui\orcidlink{0009-0006-9100-8340}}
\affiliation{RIKEN Center for Advanced Photonics, RIKEN, 2-1 Hirosawa, Wako, Saitama, 351-0198, Japan}
\affiliation{Department of Nuclear Engineering and Management, Graduate School of Engineering, The University of Tokyo, 7-3-1 Hongo, Bunkyo-ku, Tokyo 113-8656, Japan}

\author{Kenichi L. Ishikawa\orcidlink{0000-0003-2969-0212}}
\affiliation{Department of Nuclear Engineering and Management, Graduate School of Engineering, The University of Tokyo, 7-3-1 Hongo, Bunkyo-ku, Tokyo 113-8656, Japan}
\affiliation{Photon Science Center, Graduate School of Engineering, The University of Tokyo, 7-3-1 Hongo, Bunkyo-ku, Tokyo 113-8656, Japan}
\affiliation{Research Institute for Photon Science and Laser Technology, The University of Tokyo, 7-3-1 Hongo, Bunkyo-ku, Tokyo 113-0033, Japan}

\author{Yuya Morimoto\orcidlink{0000-0003-4918-2709}}
\affiliation{RIKEN Center for Advanced Photonics, RIKEN, 2-1 Hirosawa, Wako, Saitama, 351-0198, Japan}
\affiliation{RIKEN Pioneering Research Institute, RIKEN, 2-1 Hirosawa, Wako, Saitama, 351-0198, Japan}

\begin{abstract}
While the coupling of free-electron beams and light has enabled the control of electron's quantum states and the generation of ultrashort electron pulses, it can distort the measured signal when probing a laser-driven sample with an electron beam. Here, we theoretically study optical-field-driven beam deflection that inevitably occurs at the location of a laser-excited sample. Treating electrons in a beam as classical point particles and focusing on thin planar samples, we conduct a systematic study of the instantaneous deflection amplitudes with respect to light polarization, material type, thickness, and interaction geometry. For $s$-polarized excitation fields, the deflection amplitude is strictly proportional to the excitation-field amplitude in a foil, making beam deflection unavoidable. Conversely, for $p$-polarized fields, a nontrivial relationship emerges between the deflection amplitude and the field amplitude due to the interplay of electric and magnetic fields coupled with the electron--light velocity mismatch. Crucially, we demonstrate that the deflection can be minimized even under high field strengths by selecting an optimal configuration for a given foil material and thickness. These findings provide key guidelines for designing future time-resolved imaging experiments using ultrashort electron beams.
\end{abstract}
\maketitle

\section{Introduction}

Ultrafast electron microscopy and diffraction are powerful tools to investigate ultrafast structural dynamics on the nano or atomic scale~\cite{King2005JAP,Zewail2010Science,Sciaini2011RPP,Filippetto2022RMP}. Typically, a dynamical process is initiated by an ultrashort laser pulse and probed by a time-delayed electron pulse in the absence of the pump field. A crystalline foil with a thickness of tens to hundreds of nanometers is usually employed as a sample in transmission electron diffraction~\cite{Sciaini2011RPP,Morimoto2017JAP}. In contrast, due to the recent development of ultrashort electron beam technologies, dynamics occurring on timescales faster than the pump pulse duration is becoming accessible. The shaping of free-electron beams with light waves~\cite{RoadmapFreeElectrons2025ACSPhotonics,Vanacore2018NatCommun} has enabled the generation and detection of attosecond electron pulses~\cite{Priebe2017NatPhotonics,Ryabov2020SciAdv,Tsarev2021PRResearch,Morimoto2024PRL,Morimoto2018PRA,MorimotoBaum2018NatPhys,Kozak2018PRL,Tachibana2026}, thereby leading to the observation of dynamics at the sub-optical-cycle level~\cite{Gaida2024NatPhotonics,Bucher2024NatPhotonics}. The propagation and oscillation of electromagnetic fields were visualized on their natural spatial and temporal scales~\cite{Nabben2023Nature}. Modulation of transmission electron diffraction intensities on the attosecond timescale has also been reported~\cite{Morimoto2024PRL}. In these experiments, the optically modulated ultrashort electron pulses probed the sample in the presence of the excitation light field.

As exemplified by Lorentz electron microscopy, electron trajectories and
phases are highly sensitive to electric and magnetic fields, enabling
electromagnetic field imaging. Conversely, when samples under strong
electromagnetic fields are observed, interactions between free electrons and
the fields can hinder accurate observation of the sample state. Quantum
mechanically, an electron in an oscillating electromagnetic field is described
by a superposition of states involving virtual-photon absorption and emission
processes~\cite{Joachain2012Book}. In electron-beam scattering by atomic targets
under an oscillating electric field (i.e., free-free transition), the energy of the
scattered electrons changes, accompanied by a modulated angular distribution ~\cite{Kroll1976PRA,Weingartshofer1977PRL,Morimoto2015PRL,Morimoto2014JCP}.
Classical mechanically, the energy, momentum, and position of the beam undergo
time-periodic modulation due to the time-dependent Lorentz force. Attosecond electron
diffraction from a crystalline foil exposed to a laser field exhibited
time-periodic modulations in diffraction intensity, driven mainly by
field-induced instantaneous beam deflection~\cite{Morimoto2024PRL}. To
faithfully probe the structure of a sample under excitation using electron
diffraction, the field-driven shifts in the beam's momentum have to be
suppressed. On the other hand, one can actively utilize the ultrafast position
or momentum shift to, for example, slice electrons within a short time window
by transiently enhancing the diffraction probability. In order to control the
influence of the optical field on the scattering process, comprehensive
knowledge of the ultrafast light-electron interaction in the vicinity of a
sample, especially inside a foil, is necessary. However, both experimental and
theoretical studies on optically-excited foil~\cite{Kirchner2014NatPhotonics,
Ehberger2018PRL,Ehberger2019PRApplied,Feist2020PRResearch,
Ferrari2025ACSPhotonics,Morimoto2022AnnPhys,Plettner2005PRL,
Madan2022ACSPhotonics,Meuret2024ACSPhotonics,Wang2024PRB,
Muller2024Arxiv,Taleb2025NatCommun} so far have considered only the energy and
momentum shifts remaining after the completion of the electron-light
interaction, as they are directly observable quantities.

In this study, we present a systematic study of the light-field-induced transverse momentum modulation of sub-relativistic electron beams in planar foils which serve as a representative sample form. Using analytically obtained optical fields around a foil~\cite{Morimoto2018PRA}, we calculate the electromagnetic response of an electron described as a point particle. The electron transverse momentum shift is evaluated by integrating the Lorentz force along the beam trajectory. We compare the amplitudes of the transient beam deflection for different foil materials, thicknesses, light polarizations, wavelengths, and configurations. By identifying the physical mechanisms responsible for the deflection, we reveal the conditions for suppressing transient beam deflection while maintaining high electric field strength, which are necessary for future attosecond electron diffraction experiments.

The remainder of this paper is organized as follows. Section~II introduces the theoretical model and defines the quantities which characterize the instantaneous deflection. Sections~III and IV present the results for $p$-polarized and $s$-polarized light fields, respectively. Section~V summarizes the main findings.

\section{Theoretical Methods}

\begin{figure*}[t]
    \centering
    \includegraphics[width=17.2cm]{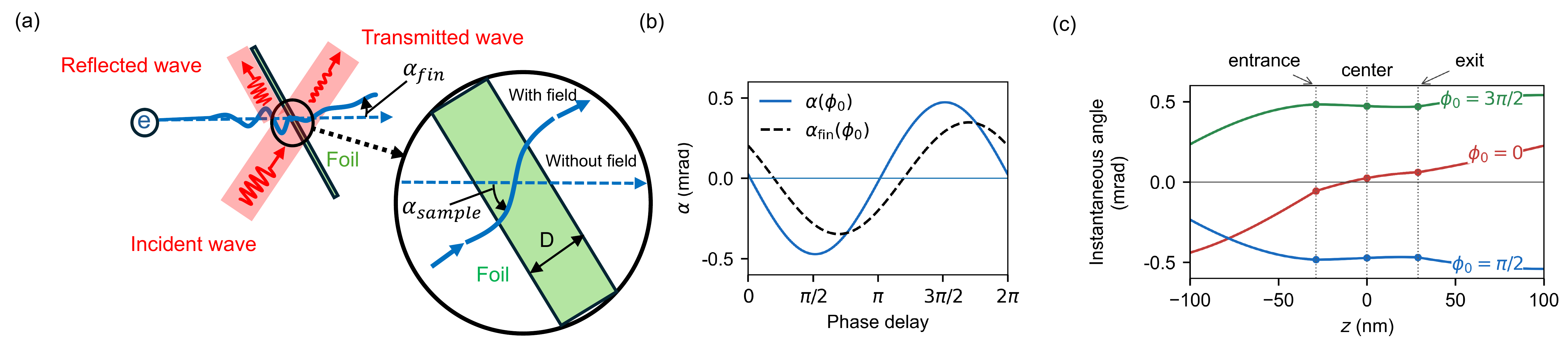}
\caption{(a) Schematic of the electron-light interaction at a planar foil. The interaction with the light field (red) modulates the electron-beam trajectory (blue) and thus induces a phase-dependent instantaneous deflection angle at the foil center, $\alpha(\phi_0)$, and a final deflection angle after the interaction, $\alpha_{\mathrm{fin}}(\phi_0)$. (b) Deflection angles as functions of the incident light phase delay $\phi_0$. The calculation is performed for a 100-keV electron interacting with a 1030-nm laser field of incident amplitude $E_0=1.0\times10^9~\mathrm{V/m}$. The sample is a 50-nm-thick Si foil. The selected geometry is $\theta_{\mathrm{laser}}=90^\circ$ and $\theta_{\mathrm{foil}}=30^\circ$, see Fig. 2(a). (c) Instantaneous electron deflection angle as a function of the longitudinal position 
$z$ for $\phi_0=\pi/2$ (blue), $\phi_0=0$ (red), and 
$\phi_0=3\pi/2$ (green). The vertical dotted lines indicate the entrance, center, and exit positions of the foil.}
    \label{fig:1}
\end{figure*}

An analytical approach is employed to investigate the momentum modulation of a swift electron [blue in Fig.~\ref{fig:1}(a)] induced by the optical field (red) inside and around a laser-excited sample foil (green rectangle). This interaction gives rise to an instantaneous deflection at the sample position, as illustrated in the inset of Fig.~\ref{fig:1}(a). Both metallic and dielectric foils are considered.

The sample foil is modeled as a laterally infinite slab with uniform thickness $D$. The incident field is linearly polarized and a monochromatic plane wave. If necessary, the deflection caused by a pulsed field can be expressed as a coherent superposition of those obtained with monochromatic plane waves. The electromagnetic field is given by the superposition of the incident, reflected, and transmitted waves, including thin-film interference. The refractive index of the foil is used as a parameter. The explicit analytical expressions of the fields and the corresponding calculation procedure are given in a previous study~\cite{Morimoto2018PRA}.

The momentum change of the electron at the foil center is obtained by integrating the Lorentz force over time along the trajectory from the remote past to the foil center,
\begin{equation}
\Delta \mathbf{p}(\phi_0)= -e \int_{-\infty}^{0}
\left[
\mathbf{E}(\mathbf{r}_e(t),t;\phi_0)
+\mathbf{v}_e \times \mathbf{B}(\mathbf{r}_e(t),t;\phi_0)
\right] \mathrm{d}t,
\end{equation}
where $e$ ($>0$) is the elementary charge and the instant at which the electron reaches the foil center is taken as $t=0$ without loss of generality. Here, $\mathbf{E}(\mathbf{r}_e(t),t;\phi_0)$ and $\mathbf{B}(\mathbf{r}_e(t),t;\phi_0)$ denote the optical electric and magnetic fields, respectively, at the position $\mathbf{r}_e(t)$. The incident electric field is expressed as $\mathbf{E}_{\mathrm{in}}(\mathbf{r},t)
=
\boldsymbol{\epsilon}_{\mathrm{in}}E_0
\cos\left(
\mathbf{k}_{\mathrm{in}}\cdot\mathbf{r}
-\omega t+\phi_0
\right)$. Here, we define $\phi_0$ as the phase delay of the incident light. The transmitted, reflected, and refracted waves are also given analytically as in Ref.~\cite{Morimoto2018PRA}. Far from the foil ($t=\pm\infty$), the field amplitude is assumed to vanish smoothly and asymptotically.

At the laser field amplitude considered in this work ($10^9~\mathrm{V/m}$, corresponding to the intensity of $1.3\times10^{11}~\mathrm{W/cm^2}$), the transverse displacement of the electron and the velocity modulation are negligible compared with its longitudinal motion and incident velocity, respectively, and therefore the electron trajectory is approximated as a straight line,
\begin{equation}
\mathbf{r}_e(t) \approx \mathbf{v}_e t,
\end{equation}
where the center of the foil is assumed to be at the origin of the spatial coordinate, and $\mathbf{v}_e=v_e\hat{\mathbf{z}}$ is the incident electron velocity. The instantaneous deflection angle is then defined as
\begin{equation}
\alpha(\phi_0) \approx \frac{\Delta p_\perp(\phi_0)}{p_{z0}},
\end{equation}
where $p_{z0}=\gamma m_e v_e$ is the initial longitudinal momentum of the incident electron, $m_e$ is the electron rest mass, $\gamma$ is the Lorentz factor, and $\Delta p_\perp(\phi_0)$ is the light-induced transverse momentum shift. The direction of the transverse deflection changes depending on the geometry and polarization. Because it takes only tens to hundreds of attoseconds for a swift electron to pass through a thin sample with a thickness of tens of nanometers, the angle can be considered nearly constant during the time the electron passes through it. Figure 1(c) shows the evolution of the instantaneous electron deflection angle along the propagation direction for three representative phase delays. Indeed, the angle remains almost constant while the electron traverses the foil.

Because the driving optical field is monochromatic and periodic in time, $\Delta p_\perp(\phi_0)$ and $\alpha(\phi_0)$ are periodic functions of $\phi_0$. We therefore scan $\phi_0$ over $2\pi$ and evaluate the amplitude of the instantaneous deflection angle,
\begin{equation}
\alpha_{\mathrm{sample}} \equiv \max_{\phi_0 \in [0,2\pi]} |\alpha(\phi_0)|.
\end{equation}
The blue curve in Fig.~\ref{fig:1}(b) shows the instantaneous deflection at the foil center, $\alpha(\phi_0)$, together with the final deflection angle at $t=+\infty$, $\alpha_{\mathrm{fin}}(\phi_0)$, shown by the black dashed line. The calculation conditions are given in the figure caption. Both $\alpha(\phi_0)$ and $\alpha_{\mathrm{fin}}(\phi_0)$ oscillate sinusoidally with a period of $2\pi$. This shows that both the instantaneous and final deflections originate from a linear interaction with the optical electromagnetic field, unlike the Kapitza-Dirac effect~\cite{Freimund2001Nature} associated with the ponderomotive effect. Because both the amplitudes and the phase offsets differ, $\alpha_{\mathrm{sample}}$ should be investigated separately from $\alpha_{\mathrm{fin}}$, highlighting the importance of separately investigating $\alpha_{\mathrm{sample}}$. Moreover, $\alpha_{\mathrm{fin}}$ vanishes in the absence of the foil, as required by energy-momentum conservation~\cite{Morimoto2018PRA}, whereas $\alpha_{\mathrm{sample}}$ can be finite even without a foil.

Throughout this work, we consider a geometric configuration in which the normal vector of the foil surface lies in the plane defined by the wave vectors of the electron and the incident light. As shown in Fig.~\ref{fig:2}(a), $\theta_{\mathrm{laser}}$ denotes the angle between the incident-light wave vector and the electron propagation direction, whereas $\theta_{\mathrm{foil}}$ denotes the angle between the foil normal and the electron propagation direction. Based on these definitions, the calculation of $\alpha_{\mathrm{sample}}$ is repeated over the angle pairs $(\theta_{\mathrm{laser}},\theta_{\mathrm{foil}})$, yielding the dependence on these angles, as shown in Figs.~\ref{fig:2}(b)--\ref{fig:2}(d). In the upper-left region of the $\alpha_{\mathrm{sample}}$ distribution, the laser and electron are incident on the foil from the same side of the foil, whereas in the lower-right region they are incident from opposite sides of the foil. The deflection amplitudes are generally larger in the upper-left region because of the smaller electron--light phase mismatch along the $z$-direction. This effect is represented by a factor proportional to $1/\Omega$ arising from the temporal integral in Eq.~(1), where $\Omega=\omega-k_{\parallel}v_{\mathrm{e}}$ and $k_{\parallel}$ is the $z$-component of the optical wave vector. Unless otherwise specified, the calculations are performed for an electron kinetic energy of 100~keV and an incident laser wavelength of 1030~nm at an amplitude of $10^9~\mathrm{V/m}$.

\section{Instantaneous deflection under $p$-polarized excitation}

\begin{figure*}[t]
    \centering
    \includegraphics[width=17.2cm]{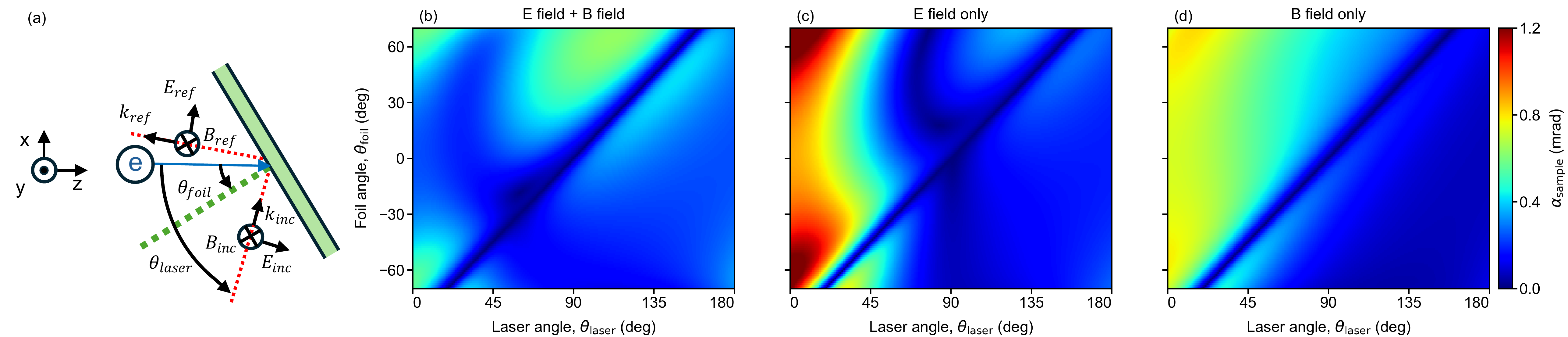}
\caption{(a) Geometry of the $p$-polarized light-electron interaction at a planar foil. The laser angle $\theta_{\mathrm{laser}}$ and the foil angle $\theta_{\mathrm{foil}}$ define the configuration. (b)--(d) Instantaneous deflection amplitude at the sample, $\alpha_{\mathrm{sample}}$, over the angles of $(\theta_{\mathrm{laser}},\theta_{\mathrm{foil}})$ for a 50-nm-thick Si foil: (b) total Lorentz-force contribution, (c) electric-field contribution only, and (d) magnetic-field contribution only.}
    \label{fig:2}
\end{figure*}

This section discusses the instantaneous deflection under $p$-polarized excitation, for which the deflection is restricted to the $x$-direction [see Fig.~\ref{fig:2}(a)]. We first investigate the relative contributions of electric and magnetic fields of light. We then compare $\alpha_{\mathrm{sample}}$ for different materials and foil thicknesses. We analyze the relationship between $\alpha_{\mathrm{sample}}$ and the electric field amplitudes in a foil and discuss geometries suitable for sub-cycle electron diffraction. We introduce a simple model for qualitatively predicting $\alpha_{\mathrm{sample}}$ without elaborate calculations. We then discuss the dependences on electron kinetic energy and laser wavelength.

\begin{figure*}[t]
    \centering
    \includegraphics[width=17.2cm]{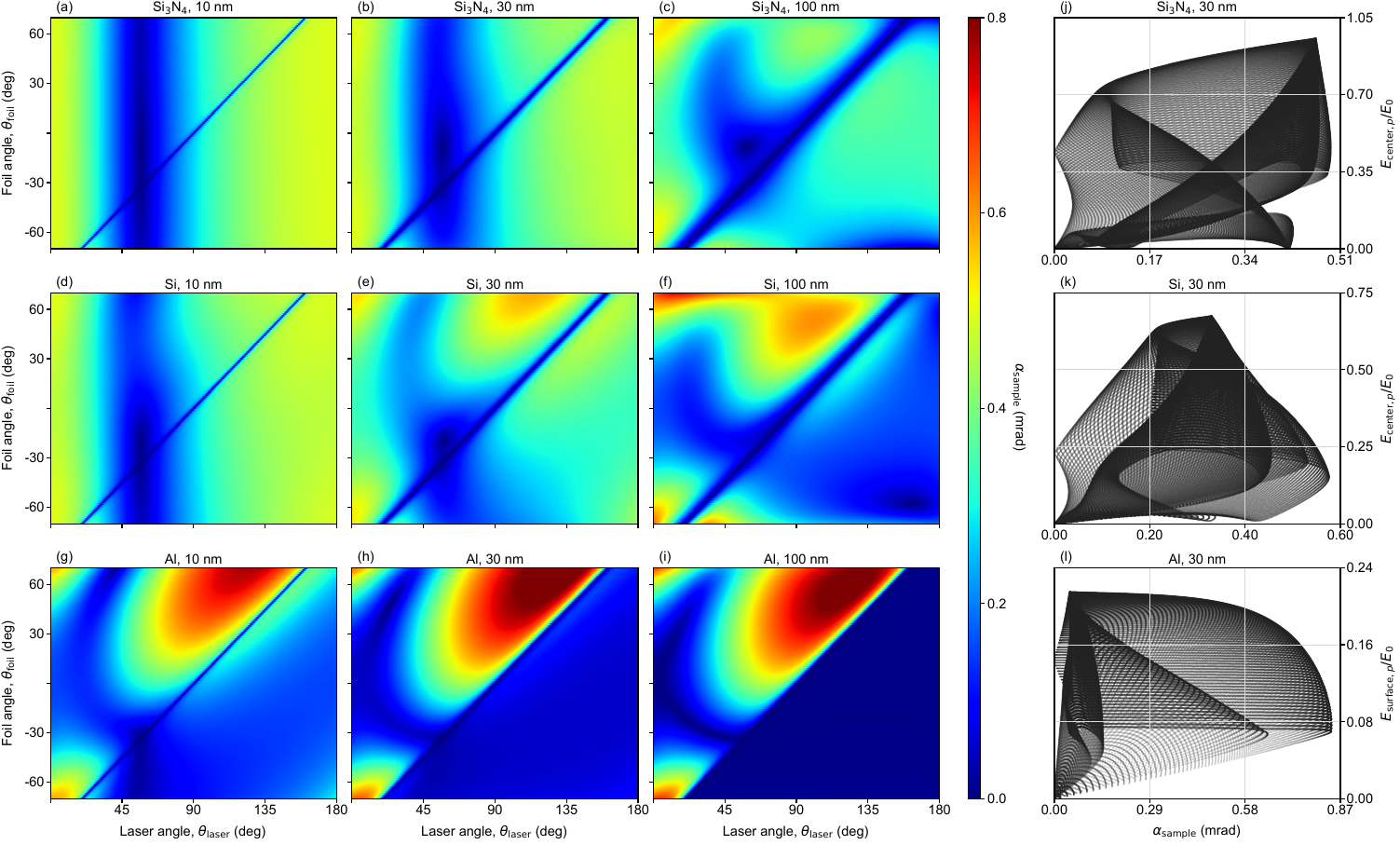}
\caption{Instantaneous deflection amplitudes at the planar sample, $\alpha_{\mathrm{sample}}$, under $p$-polarized excitation for different foil materials and thicknesses. Panels (a)--(i) show $\alpha_{\mathrm{sample}}$ for each $(\theta_{\mathrm{laser}},\theta_{\mathrm{foil}})$. The rows correspond to Si$_3$N$_4$, Si, and Al, and the columns to thicknesses of 10, 30, and 100~nm. Panels (j) and (k) show the relation between the normalized foil-center electric-field amplitude, $E_{\mathrm{center},p}/E_0$, and $\alpha_{\mathrm{sample}}$ for 30-nm-thick Si$_3$N$_4$ and Si foils, respectively. Panel (l) shows the ratio $E_{\mathrm{surface},p}/E_0$ for 30-nm-thick Al foils. Each point in panels (j)--(l) corresponds to one angular pair $(\theta_{\mathrm{laser}},\theta_{\mathrm{foil}})$.}
    \label{fig:3}
\end{figure*}

\subsection{Importance of Magnetic Field Contribution}

At the kinetic energy of 100~keV considered in this work, the velocity of the free electrons reaches 54.8\% of the speed of light. Therefore, the magnetic term in the Lorentz force [Eq.~(1)] cannot be neglected. To analyze its contribution, we consider a 50-nm-thick Si foil as an example. Figure~\ref{fig:2}(b) shows the $\alpha_{\mathrm{sample}}$ obtained from the combined action of the electric- and magnetic-field terms, while Figs.~\ref{fig:2}(c) and \ref{fig:2}(d) show the corresponding results obtained by retaining only the electric- and the magnetic-field terms, respectively. Although these contributions are not independently observable, their comparison shows that the magnetic-field contribution is not negligible and can significantly modify the total deflection. The fine structures appearing in Fig.~2(b) mainly originate from the electric-field contribution [Fig.~2(c)]. For typical dielectrics, the $x$-components of the incident and reflected electric fields have opposite signs in the vicinity of the foil [see Fig.~2(a)], and therefore interfere destructively. By contrast, the magnetic fields of the incident and reflected waves have the same sign and therefore interfere constructively.

\subsection{Material and Thickness Dependence}

We compare $\alpha_{\mathrm{sample}}$ for three representative foil materials at three different thicknesses. As a dielectric with a relatively low refractive index, we choose silicon nitride (Si$_3$N$_4$, $n=2.01$), which is commonly used as a sample-support membrane in electron microscopy. For comparison, we also consider silicon (Si, $n=3.56$) as a dielectric with a higher refractive index, and aluminum (Al, $n+ik=1.40+9.85i$) as a metallic foil. For each material, we consider three different thicknesses within the range commonly used in transmission electron microscopy: 10, 30, and 100~nm.

Figure~\ref{fig:3} shows the calculation results. The most pronounced difference is observed between the metallic and dielectric foils. Al (bottom row) exhibits a larger overall deflection amplitude than Si (middle row) and Si$_3$N$_4$ (top row) for the same thickness. In contrast, Si and Si$_3$N$_4$ show a high degree of similarity, with $\alpha_{\mathrm{sample}}$ remaining below about 0.5~mrad over most configurations.

We examine the thickness dependence of Si$_3$N$_4$. In Fig.~\ref{fig:3}(a), the low-deflection region appears as a broad vertical stripe extending over $\theta_{\mathrm{laser}}\approx30^\circ$--$90^\circ$. This nearly vertical feature indicates a weak dependence on $\theta_{\mathrm{foil}}$ and suggests that the contribution of the reflected wave is negligibly small for the 10-nm-thick Si$_3$N$_4$ foil. In Fig.~\ref{fig:3}(b), and more clearly in Fig.~\ref{fig:3}(c), this low-deflection feature becomes inclined and extends over a much broader range of $\theta_{\mathrm{laser}}$, approximately from $25^\circ$ to $100^\circ$. At the same time, the high-deflection region becomes more concentrated at the corners.

A similar but stronger thickness dependence is seen in Figs.~\ref{fig:3}(d)--\ref{fig:3}(f) for Si. A substantial change appears in Fig.~\ref{fig:3}(f), where the low-deflection feature becomes much more inclined than that in Si$_3$N$_4$, and the high-deflection regions become more evident in the left corners. This comparison shows that the thickness dependence is much stronger for Si than for Si$_3$N$_4$ in the 10--30~nm range, whereas the 100-nm cases already differ substantially from the thinner foils for both dielectrics.

By contrast, for Al, the patterns do not strongly depend on the thickness. In particular, those at 30 and 100~nm are nearly identical [Figs.~\ref{fig:3}(h)--\ref{fig:3}(i)]. The optical penetration depth of Al is approximately 17~nm. Therefore, once the foil thickness reaches a few tens of nanometers, the optical fields inside and behind the foil are strongly attenuated, and the response becomes dominated by surface reflection. This also explains the low-deflection region at the lower right, where the electron interacts mainly with the attenuated transmitted field because the light is incident from the opposite side of the foil. In the limit of total reflection, $\alpha_{\mathrm{sample}}$ coincides with $\alpha_{\mathrm{fin}}$, and the configuration for a vanishing deflection angle is given by electron--light velocity matching along the $x$-direction~\cite{Morimoto2018PRA,Ehberger2018PRL}. The above results show that, under $p$-polarized excitation, the magnitude and configuration dependence of $\alpha_{\mathrm{sample}}$ vary with the foil material and thickness, while the strength of the thickness dependence is material dependent.

\subsection{Correlation with the Electric-Field Amplitude at the Foil Center}

From an experimental perspective, the electric-field amplitude at the foil center is also of considerable interest. To investigate the relationship between $\alpha_{\mathrm{sample}}$ and the electric-field amplitude inside the foil under $p$-polarized excitation, we define
\begin{equation}
E_{\mathrm{center},p}
\equiv
\max_{\phi_0\in[0,2\pi]}
\left|
E_{x,p}(\mathbf{r}_{\mathrm{center}};\phi_0)
\right|,
\end{equation}
where $E_{x,p}$ denotes the $x$-component of the electric field under $p$-polarized excitation, and $\mathbf{r}_{\mathrm{center}}=0$ denotes the position at the center of the foil. Note that the value of $\phi_0$ that maximizes $|E_{x,p}(\mathbf{r}_{\mathrm{center}};\phi_0)|$ does not necessarily coincide with that which maximizes $|\alpha(\phi_0)|$. If a nontrivial relationship exists between $\alpha_{\mathrm{sample}}$ and $E_{\mathrm{center},p}$, configurations may be identified in which the sample is strongly excited while the electron-beam direction is only weakly modulated. To examine this relationship, we plot $\alpha_{\mathrm{sample}}$ against $E_{\mathrm{center},p}/E_0$ in Figs.~\ref{fig:3}(j) and \ref{fig:3}(k), where $E_0$ is the incident optical-field amplitude. Each point corresponds to a geometric configuration $(\theta_{\mathrm{laser}},\theta_{\mathrm{foil}})$.

In Fig.~\ref{fig:3}(j), the data points for the 30-nm-thick Si$_3$N$_4$ foil are distributed over approximately 0--0.50~mrad along the horizontal axis ($\alpha_{\mathrm{sample}}$) and 0--1.0 along the vertical axis ($E_{\mathrm{center},p}/E_0$). The distribution is broad, with multiple different values of $E_{\mathrm{center},p}/E_0$ corresponding to similar values of $\alpha_{\mathrm{sample}}$. Similarly, no clear one-to-one correspondence is found for the 30-nm-thick Si [Fig.~\ref{fig:3}(k)] and 30-nm-thick Al [Fig.~\ref{fig:3}(l)] foils. Because the optical field is strongly attenuated in Al foils, we evaluate the electric field amplitude at the light-incident surface ($E_{\mathrm{surface},p}$) rather than at the center. These results show that the two quantities do not exhibit a simple linear relation under $p$-polarized excitation. The main reason is that $\alpha_{\mathrm{sample}}$ is not a local quantity, but an accumulated response of the electron to the optical field along its trajectory.

To identify experimentally favorable configurations, we searched for angular points that simultaneously give a small instantaneous deflection ($\alpha_{\mathrm{sample}}<0.1~\mathrm{mrad}$) and a large electric-field amplitude at the foil center ($E_{\mathrm{center},p}/E_0>0.5$). For the 30-nm-thick Si$_3$N$_4$ foil, this condition is satisfied around $\theta_{\mathrm{laser}}\simeq 48^\circ$--$51^\circ$ and $\theta_{\mathrm{foil}}\simeq 15^\circ$--$27^\circ$, where $E_{\mathrm{center},p}/E_0$ reaches approximately 0.75. For the 30-nm-thick Si foil, no point in the present angular scan meets this criterion; relaxing the criterion to $\alpha_{\mathrm{sample}}<0.15$~mrad yields favorable points around $\theta_{\mathrm{laser}}\simeq 43^\circ$--$51^\circ$ and $\theta_{\mathrm{foil}}\simeq -12^\circ$ to $-4^\circ$, with $E_{\mathrm{center},p}/E_0\simeq 0.50$--$0.52$. Although the available foil angles $\theta_{\mathrm{foil}}$ are limited when observing the transmission diffraction of crystalline samples, the deflection amplitudes can be controlled by adjusting $\theta_{\mathrm{laser}}$ and the foil thickness.

\begin{figure}[t]
    \centering
    \includegraphics[width=\columnwidth]{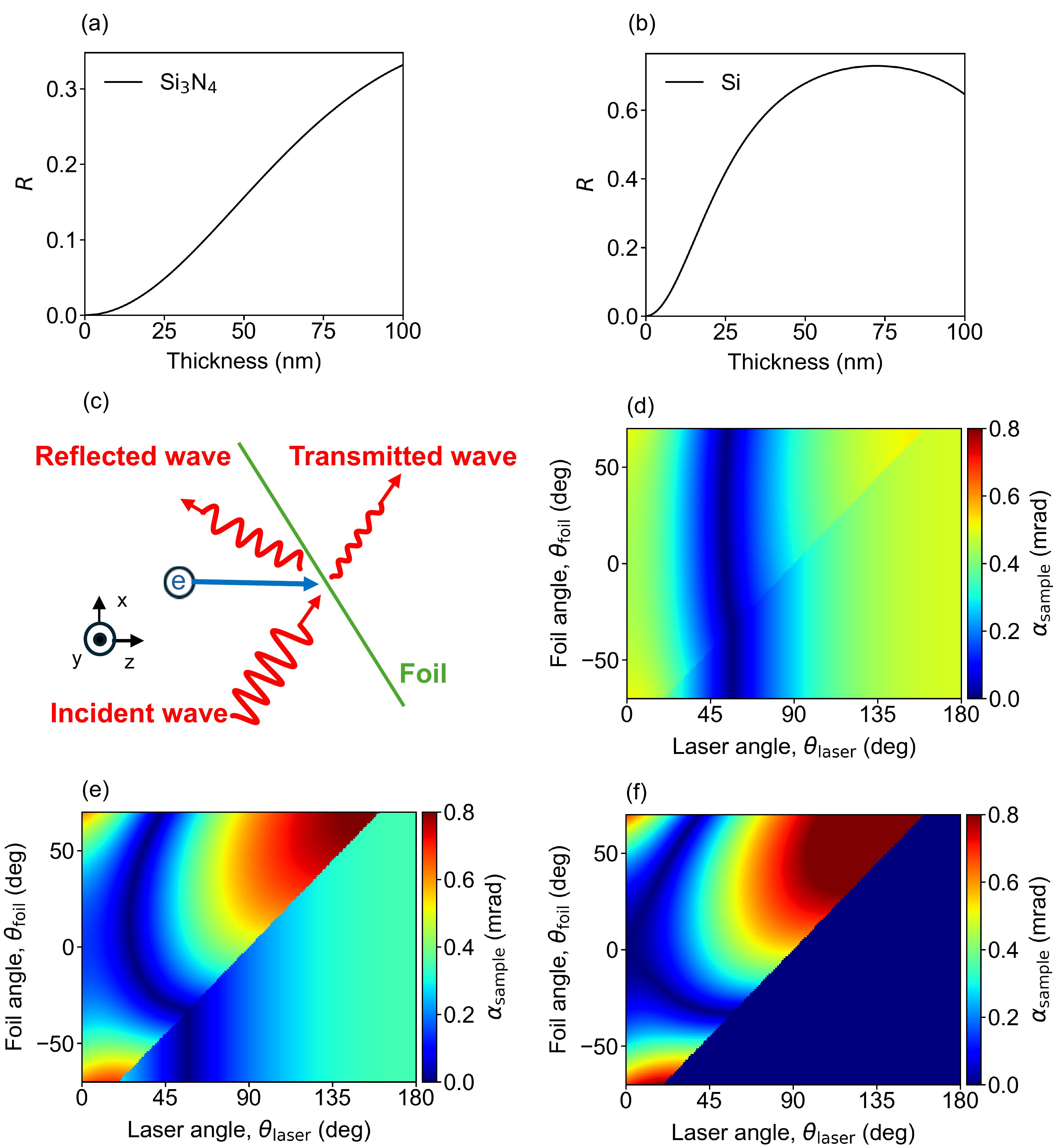}
\caption{Three-wave model for the instantaneous deflection under $p$-polarized excitation. (a) and (b) Normal-incidence reflectivities $R$ as a function of foil thickness for Si$_3$N$_4$ and Si, respectively. (c) Schematic of the simplified geometry, where the foil is treated as a zero-thickness interface and only the incident, reflected, and transmitted waves are considered. (d)--(f) Calculated $\alpha_{\mathrm{sample}}$ for $R=0.01$ (d), 0.5 (e), and 1.0 (f).}
    \label{fig:4}
\end{figure}

\subsection{Three-Wave Model: A Simplified Picture}

To extract the physical mechanism underlying the thickness and
configuration dependence of $\alpha_{\mathrm{sample}}$ discussed in
Sec.~III B, we introduce a simplified three-wave model. In this model,
we neglect the light--electron interaction inside the foil, and the foil
is approximated as a zero-thickness interface crossing at $x$=$z$=0. In
addition, only three plane-wave components are retained: the incident,
reflected, and transmitted waves, as shown in Fig.~\ref{fig:4}(c). The electromagnetic fields considered for the Lorentz-force in Eq.~(1) are chosen according
to the side from which the light reaches the interface. In the upper-left
region of the configuration-dependence plots, where the laser and electron
are incident from the same side of the foil, the electron interacts with
the superposition of the incident and reflected waves before reaching the
foil. In the lower-right region, where the laser is incident from the
opposite side, the electron interacts only with the transmitted wave.

The reflection and transmission coefficients for $p$-polarized
light denoted by $r_p$ and $t_p$ are expressed by $r_p=-|r|$ and $t_p=\sqrt{1-|r|^2}$, where a reflection phase shift of
$\pi$ and a transmission phase shift of zero are assumed. The corresponding
power reflectivity is $R=|r_p|^2=|r|^2$. Within this framework, the
complicated material- and thickness-dependent optical responses are
represented by the single parameter $R$.

In this model, the deflection
amplitude is analytically given by
\begin{equation}
\alpha_{\mathrm{sample}}
=\frac{eE_0}{\omega p_{z0}}
\begin{cases}
\left|g_{\mathrm i}+\sqrt{R}\,g_{\mathrm r}\right|,
& \text{same-side incidence},\\[3pt]
\sqrt{1-R}\,\left|g_{\mathrm i}\right|,
& \text{opposite-side incidence},
\end{cases}
\end{equation}
where $g(\theta)=(\cos\theta-\beta)/(1-\beta\cos\theta)$,
$\beta=v_e/c$, $g_{\mathrm i}=g(\theta_{\mathrm{laser}})$ for the
incident wave, and
$g_{\mathrm r}=g(\pi+2\theta_{\mathrm{foil}}
-\theta_{\mathrm{laser}})$ for the reflected wave.
Here, same-side and opposite-side incidence refer to configurations
in which the laser and electron are incident from the same and opposite
sides of the foil, respectively. Thus, the model directly connects
$\alpha_{\mathrm{sample}}$ to the reflectivity and geometry.

We consider three representative values: $R=0.01$, 0.5, and 1.0. The cases
$R=0.01$ and $R=0.5$ approximately represent 10-nm-thick Si$_3$N$_4$ and 30-nm-thick Si foils under normal incidence [Figs.~\ref{fig:4}(a) and
\ref{fig:4}(b)], respectively, while $R=1.0$ corresponds to a perfect
electric conductor (PEC). Figures~\ref{fig:4}(d)--\ref{fig:4}(f) show the configuration dependence of $\alpha_{\mathrm{sample}}$ for $R=0.01$, 0.5, and 1.0,
respectively. All three results share a common feature: a low-deflection
region appears in the intermediate $\theta_{\mathrm{laser}}$ range,
whereas larger deflection amplitudes appear at both smaller and larger
$\theta_{\mathrm{laser}}$. The contrast between the low- and
high-deflection regions becomes stronger as $R$ increases.

For $R=0.01$ [Fig.~\ref{fig:4}(d)], the three-wave model reproduces the
relatively weak contrast of the 10-nm-thick Si$_3$N$_4$ result
[Fig.~\ref{fig:3}(a)]: the low-deflection region extends over an
intermediate $\theta_{\mathrm{laser}}$ range, with weak dependence of $\theta_{\mathrm{foil}}$. For $R=0.5$
[Fig.~\ref{fig:4}(e)], comparison with the 30-nm-thick Si
[Fig.~\ref{fig:3}(e)] shows that overall features in the upper-left region are well reproduced,
suggesting that the dominant contribution is interference between the
incident and reflected waves. However, the lower-right region differs:
the calculation for Si shows a smoother, tilted structure, whereas the
three-wave model gives a nearly vertical stripe pattern because the field
discontinuities at the surface and the fields inside the foil are
neglected. For $R=1.0$ [Fig.~\ref{fig:4}(f)], corresponding to total
reflection, the distribution closely resembles the 30-nm-thick Al result
[Fig.~\ref{fig:3}(h)]: both show a strong-deflection region in the
upper-left and a nearly vanishing $\alpha_{\mathrm{sample}}$ in the
lower-right. The good qualitative reproducibility of the three-wave model indicates
that optical reflectivity and the associated interference between the
incident and reflected waves capture the dominant features of the
configuration dependence of $\alpha_{\mathrm{sample}}$. The model
therefore provides a simple physical picture of the instantaneous
deflection around a thin foil.

\subsection{Electron-Energy and Laser-Wavelength Dependence}

\begin{figure*}[t]
    \centering
    \includegraphics[width=12.9cm]{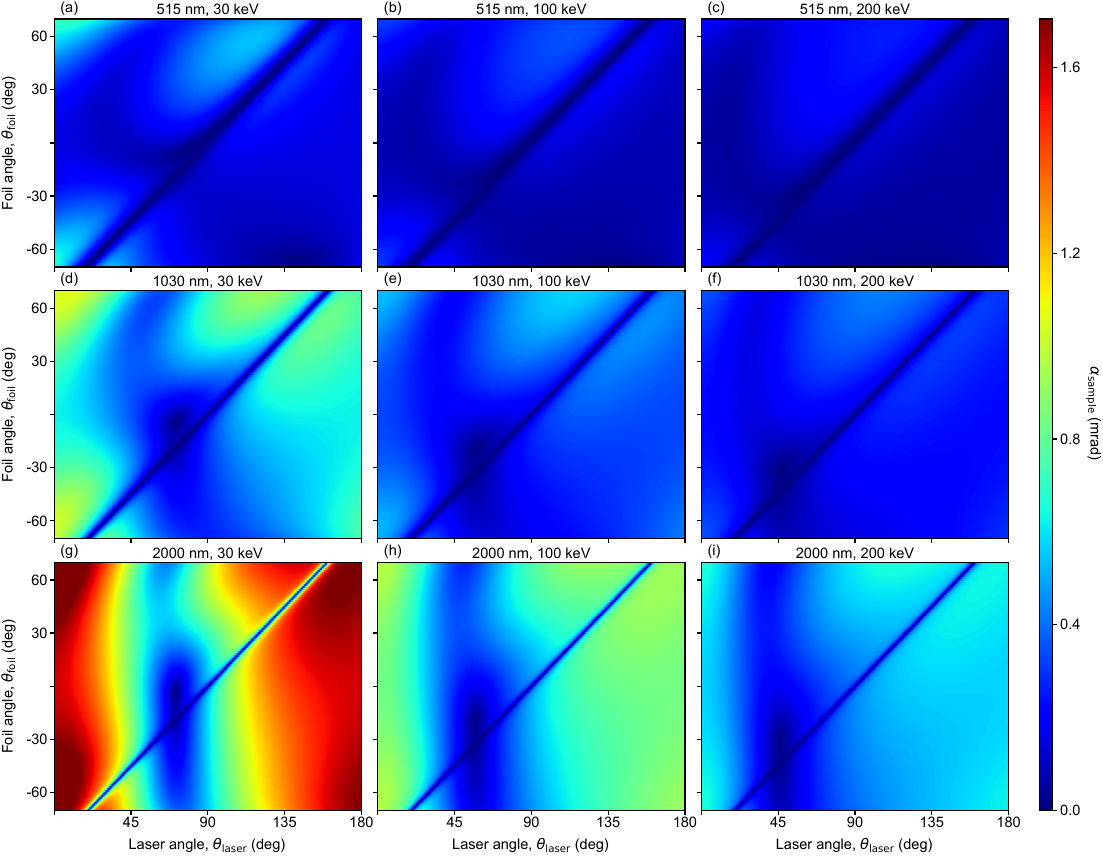}
\caption{Instantaneous deflection amplitudes, $\alpha_{\mathrm{sample}}$, for a 30-nm-thick Si foil under $p$-polarized excitation at different electron kinetic energies and laser wavelengths. Panels (a)--(c) show the results at a wavelength of 515~nm for electron kinetic energies of 30, 100, and 200~keV, respectively. Panels (d)--(f) show the corresponding results at a wavelength of 1030~nm. Panels (g)--(i) show the corresponding results at a wavelength of 2000~nm.}
    \label{fig:5_energy}
\end{figure*}

The above results are for 100-keV electron energy and 1030-nm laser wavelength. To examine the dependence of $\alpha_{\mathrm{sample}}$ on the electron energy and laser wavelength, we compare $\alpha_{\mathrm{sample}}$ for a 30-nm-thick Si foil at three different electron kinetic energies, 30, 100, and 200~keV, and at three laser wavelengths, 515~nm [Fig.~\ref{fig:5_energy}(a)--\ref{fig:5_energy}(c)], 1030~nm [Fig.~\ref{fig:5_energy}(d)--\ref{fig:5_energy}(f)], and 2000~nm [Fig.~\ref{fig:5_energy}(g)--\ref{fig:5_energy}(i)].

For a fixed laser wavelength, the overall magnitude of $\alpha_{\mathrm{sample}}$ decreases as the electron kinetic energy increases. This trend can be explained by Eqs.~(1) and (3). The transverse momentum exchange comprises electric- and magnetic-field contributions. The electric-field contribution has no direct velocity factor, whereas the magnetic-field contribution contains $v_e$ through the $\mathbf{v}_e\times\mathbf{B}$ term. However, this velocity dependence is insufficient to compensate for the increase in the initial longitudinal momentum $p_{z0}$. Therefore, $\alpha_{\mathrm{sample}}$ becomes smaller at higher electron energies.

For a fixed electron energy, the deflection generally increases as the laser wavelength is increased. This dependence can be directly understood from Eq.~(1). For a fixed incident electric-field amplitude, the time integral of an optical-cycle force gives a characteristic factor of $1/\Omega=1/(\omega-k_{\parallel}v_e)$. Therefore, a longer wavelength (i.e., a smaller $\omega$) leads to a larger momentum change and hence a stronger instantaneous deflection.

\section{Instantaneous deflection under $s$-polarized excitation}

We now turn to the case of an $s$-polarized excitation field. In contrast to the $p$-polarized case, 
where the deflection occurs along the $x$-direction, 
the deflection under $s$-polarized excitation occurs along the $y$-direction. In addition, $s$-polarized excitation does not induce any longitudinal energy exchange. Moreover, the final deflection $\alpha_{\mathrm{fin}}$ vanishes~\cite{Morimoto2018PRA}.
Therefore, the instantaneous deflection by the $s$-polarized field is very similar to that without a foil.

\subsection{Magnetic Field Contribution}

\begin{figure*}[t]
\centering
\includegraphics[width=17.2cm]{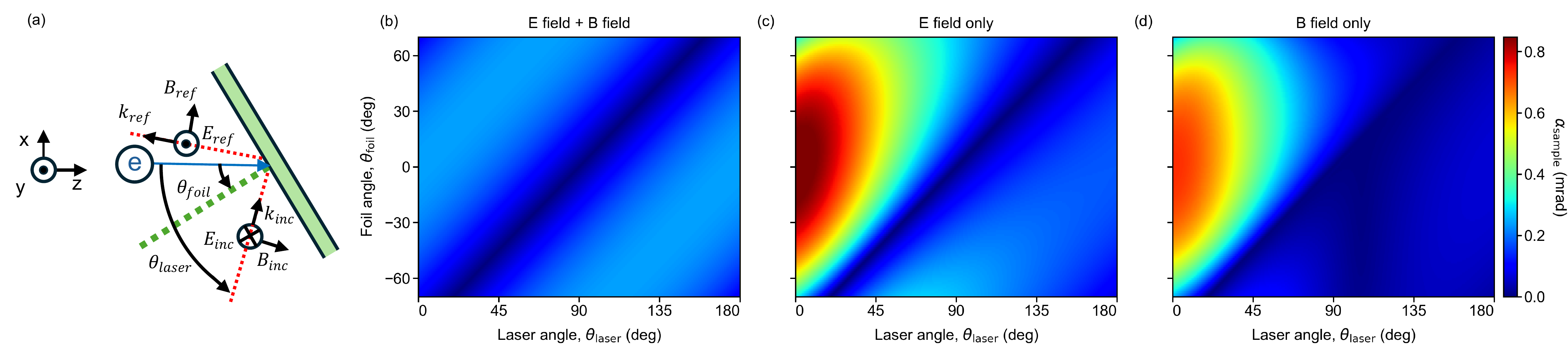}
\caption{Electric- and magnetic-field contributions to $\alpha_{\mathrm{sample}}$ under $s$-polarized excitation for a 50-nm-thick Si foil. The same set of parameters is used as in Fig.~\ref{fig:2}. (a) Illustration of the geometry. (b)--(d) Calculated $\alpha_{\mathrm{sample}}$ by considering (b) the total Lorentz-force contribution, (c) the electric-field contribution only, and (d) the magnetic-field contribution only.}
\label{fig:s_pol_EB}
\end{figure*}

We investigate the role of the magnetic field under $s$-polarized excitation by adopting the same foil thickness (50-nm-thick Si) as in the $p$-polarized case discussed in Sec.~III~A. Figures~\ref{fig:s_pol_EB}(b)--\ref{fig:s_pol_EB}(d) show $\alpha_{\mathrm{sample}}$ obtained from the combined electric and magnetic fields, the electric-field-only contribution, and the magnetic-field-only contribution, respectively. Compare to case of $p$-polarization, the dependence on the geometrical angles is much weaker because the electric-field vector points along the $y$-axis. Moreover, unlike the case of $p$-polarization, the electric and magnetic fields contribute almost equally in the upper-left angular region, and the total deflection is greatly suppressed. Due to the boundary conditions [see Fig.6(a)], the effect of the magnetic field is enhanced, whereas that of the electric field is attenuated [see Fig.~\ref{fig:s_pol_EB}(a)]. In addition, the electric and magnetic components of the Lorentz force act in opposite transverse directions. Thus partial cancellation occurs.

\subsection{Material and Thickness Dependence}
\begin{figure*}[t]
    \centering
    \includegraphics[width=17.2cm]{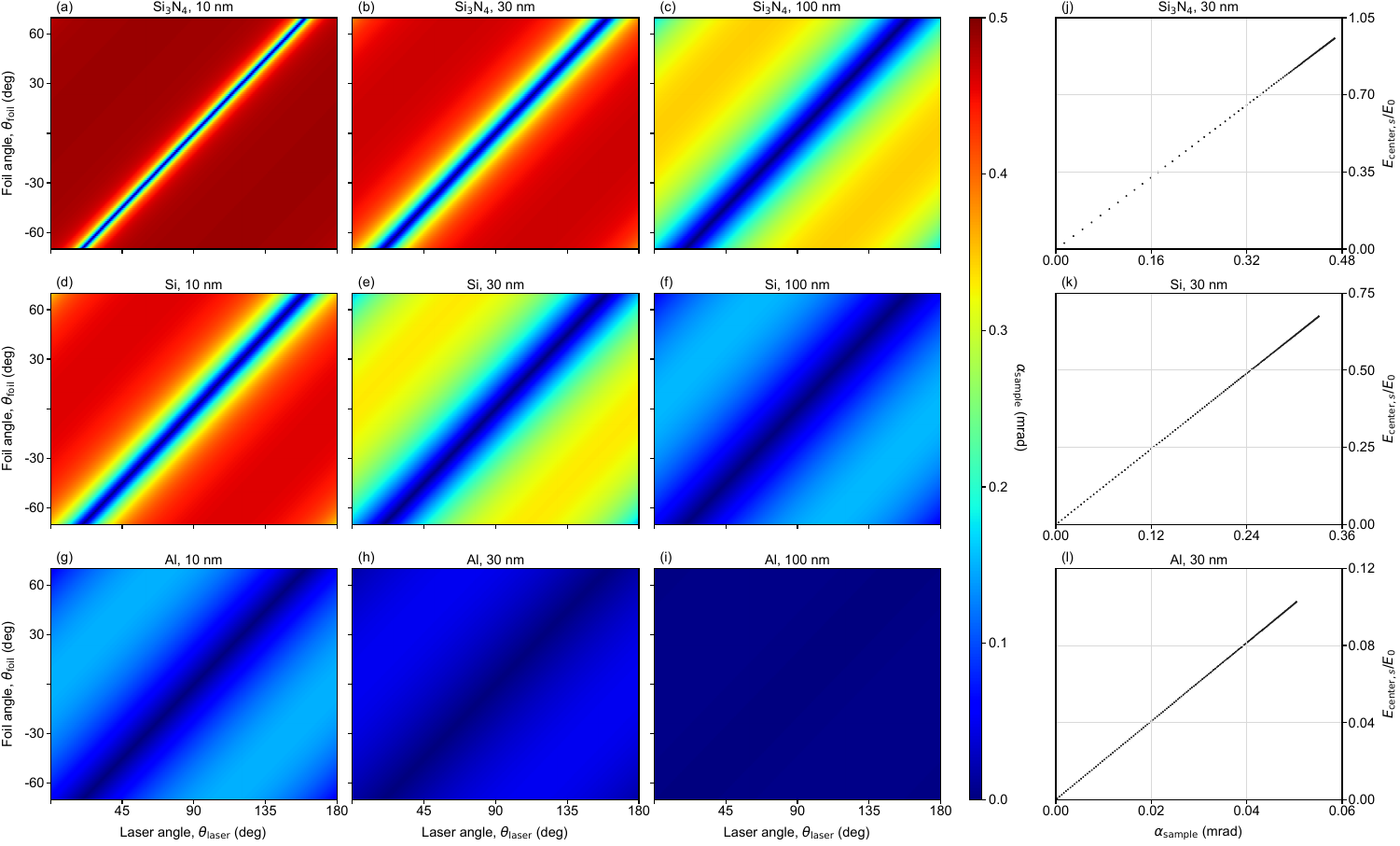}
\caption{Instantaneous deflection amplitudes at the planar sample, $\alpha_{\mathrm{sample}}$, under $s$-polarized excitation for different foil materials and thicknesses. Panels (a)--(i) show $\alpha_{\mathrm{sample}}$ for each $(\theta_{\mathrm{laser}},\theta_{\mathrm{foil}})$. The rows correspond to Si$_3$N$_4$, Si, and Al, and the columns to thicknesses of 10, 30, and 100~nm. Panels (j)--(l) show the relationship between the normalized internal $s$-polarized electric-field amplitude at the foil center, $E_{\mathrm{center},s}/E_0$, and $\alpha_{\mathrm{sample}}$ for 30-nm-thick Si$_3$N$_4$, Si, and Al foils, respectively. Each point in panels (j)--(l) corresponds to one angular pair $(\theta_{\mathrm{laser}},\theta_{\mathrm{foil}})$.}
    \label{fig:8_s_maps}
\end{figure*}

Figure~\ref{fig:8_s_maps} shows simulated $\alpha_{\mathrm{sample}}$ under $s$-polarized excitation for the three materials and three thicknesses. The overall material dependence is markedly different from that in the $p$-polarized case [Figs.~\ref{fig:3}(a)--\ref{fig:3}(i)]. For Al [Figs.~\ref{fig:8_s_maps}(g)--\ref{fig:8_s_maps}(i)], maximum $\alpha_{\mathrm{sample}}$ is approximately $0.15$~mrad at 10~nm and decreases to nearly zero at thicknesses of 30 and 100~nm. As we will see in Sec.~IV C, for $s$-polarized excitation, the instantaneous deflection amplitude at the sample position is proportional to the internal electric-field amplitude at the foil center, $E_{\mathrm{center},s}$, as in the case without a foil.

In contrast, dielectric foils exhibit nonzero deflection because a finite electric field exists inside the foil. For Si$_3$N$_4$, the 10-nm-thick result [Fig.~\ref{fig:8_s_maps}(a)] shows high deflection amplitudes over most of the angle combinations, with a narrow low-deflection band running diagonally across the map. This band satisfying $|\theta_{\mathrm{laser}}-\theta_{\mathrm{foil}}|\approx90^\circ$ corresponds to grazing incidence at the foil surface, where the incident light is almost completely reflected at the foil interface ($t_s\approx0$)~\cite{Hecht2016Optics}, resulting in a negligibly small electric field and deflection at the sample position. The weak dependence on $(\theta_{\mathrm{laser}},\theta_{\mathrm{foil}})$
stems from the nearly constant reflectivity for $s$-polarized light except grazing incidence. As the thickness increases to 30~nm [Fig.~\ref{fig:8_s_maps}(b)], the high-deflection region shrinks and the low-deflection diagonal band broadens. At 100~nm [Fig.~\ref{fig:8_s_maps}(c)], the distribution is dominated by lower values over most of the angular range. For Si, the thickness dependence follows a similar but more pronounced trend. The deflection angle decreases with increasing foil thickness.

The thickness dependence can be understood from the normal-incidence reflectivity shown in Figs.~\ref{fig:4}(a) and \ref{fig:4}(b). For $s$-polarized excitation, $\alpha_{\mathrm{sample}}$ is proportional to $E_{\mathrm{center},s}$, as will be shown in the following subsection. 
In the present thin-film geometry, a larger reflectivity  
means weaker coupling of the incident light into the foil, leading to a smaller field at the sample center and hence a smaller $\alpha_{\mathrm{sample}}$. The generally higher reflectivity of Si than that of Si$_3$N$_4$ also explains why the deflection is more strongly suppressed in Si.

The results for the dielectrics [Figs.~\ref{fig:8_s_maps}(a)--\ref{fig:8_s_maps}(f)] show an approximate symmetry between the upper-left and lower-right regions. This symmetry can also be understood based on the internal electric field. For the paired angular configurations in these two regions, the absolute angle of incidence with respect to the foil normal is the same. Therefore, the magnitude of the $s$-polarized electric field inside the foil is the same.

\subsection{Correlation with the Electric-Field Amplitude at the Foil Center}

We compare $\alpha_{\mathrm{sample}}$ with the normalized amplitude of the internal electric field, $E_{\mathrm{center},s}/E_0$. Interestingly, in strong contrast to the $p$-polarization case [Fig.~\ref{fig:3}(j)--(l)], the numerical results in Figs.~\ref{fig:8_s_maps}(j)--\ref{fig:8_s_maps}(l) show that the data points for each material collapse onto a straight line.

This linear relation can be directly understood from Eq.~(1). Let us consider an arbitrary $s$-polarized plane-wave component, $\mathbf{E}_j=(0,E_{j,y},0)$, $\mathbf{k}_j=(k_{j,x},0,k_{j,z})$, and $\mathbf{B}_j=\mathbf{k}_j\times\mathbf{E}_j/\omega
=(B_{j,x},0,B_{j,z})$, where
$B_{j,x}=-(k_{j,z}/\omega)E_{j,y}$ and
$B_{j,z}=(k_{j,x}/\omega)E_{j,y}$. The transverse Lorentz force for the electron moving along the $z$-direction is $E_{j,y}+v_eB_{j,x}=(1-v_ek_{j,z}/\omega)E_{j,y}=\Omega_j/\omega E_{j,y}$, with $\Omega_j=\omega-k_{j,z}v_e$. Since the field sampled by the electron along its trajectory $z=v_et$ oscillates as $\cos(\Omega_j t)$, the time integral in Eq.~(1) gives a factor proportional to $1/\Omega_j$, which cancels the $\Omega_j$ factor in the transverse Lorentz force. Accordingly, the temporal integral of the transverse force is independent of both the electron velocity and the electron-laser angle, and is directly connected to the electric field amplitude. 

Summing over all the plane-wave components including those inside a foil, the remaining contribution to the momentum shift is determined by the internal electric field amplitude at the foil center, giving
\begin{equation}
\alpha_{\mathrm{sample}}
=
\frac{e}{\omega p_{z0}}\,E_{\mathrm{center},s},
\label{eq:s_pol_linear}
\end{equation}
which explains the linear dependence between $\alpha_{\mathrm{sample}}$ and the internal $s$-polarized field amplitude observed in Fig.~\ref{fig:8_s_maps}(j)--\ref{fig:8_s_maps}(l).

This observation has an important implication from the experimental point of view that increasing the internal electric-field amplitude inevitably increases the instantaneous deflection angle. Thus, it is difficult to realize a strong excitation field at the sample position while keeping $\alpha_{\mathrm{sample}}$ small in the $s$-polarized geometry. Conversely, if this quantity can be translated into a measurable physical quantity, it would be possible to perform in-situ characterization of the electric field amplitudes inside a material. Equation~\eqref{eq:s_pol_linear} also indicates that, 
as in the case of $p$-polarization, a longer wavelength induces a stronger instantaneous deflection.

\section{Conclusion}

We have presented a comprehensive theoretical study of the field-driven deflection of sub-relativistic electrons within laser-excited planar foils. The amplitude of the instantaneous angle, $\alpha_{\mathrm{sample}}$, depends on light polarization, field amplitude, foil material, thickness, and interaction geometry. The magnetic field of light significantly influences the deflection regardless of the polarization direction. A key factor underlying the geometric dependence is optical reflectance and transmittance. For $p$-polarized excitation, there are nontrivial relationships between $\alpha_{\mathrm{sample}}$ and the electric field amplitude in a foil due to the complex interplay of the forces induced by the electric and magnetic fields in combination with the velocity mismatch along $z$-direction (i.e., $\Omega$). Therefore, for a foil with a given material type and thickness, there exist geometries that give nearly vanishing $\alpha_{\mathrm{sample}}$. 
In contrast, for $s$-polarized excitation, $\alpha_{\mathrm{sample}}$ scales linearly with the internal electric-field amplitude as if the foil were absent. These findings provide practical guidelines for selecting optimal experimental configurations that minimize or maximize beam deflection while maintaining strong optical excitation within the foils.

\begin{samepage}
\begin{acknowledgments}
This research was supported by MEXT/JSPS KAKENHI (Grant Nos.~JP25K22230 and JP25K01734), JST FOREST (Grant No.~JPMJFR2228), and the Gordon and Betty Moore Foundation. This work was also partially supported by the RIKEN TRIP initiative (HIKARI-COOL Tokyo). X.G. acknowledges support from the Doctoral Student Special Incentives Program for International Students (SEUT-RA Type IS), Graduate School of Engineering, The University of Tokyo.
\end{acknowledgments}

\end{samepage}

\clearpage
\bibliography{refs}

@article{King2005JAP,
  author  = {King, W. E. and Campbell, G. H. and Frank, A. and Reed, B. and Schmerge, J. F. and Siwick, B. J. and Stuart, B. C. and Weber, P. M.},
  title   = {Ultrafast electron microscopy in materials science, biology, and chemistry},
  journal = {Journal of Applied Physics},
  volume  = {97},
  pages   = {111101},
  year    = {2005},
  doi     = {10.1063/1.1927699}
}

@article{Zewail2010Science,
  author  = {Zewail, Ahmed H.},
  title   = {Four-Dimensional Electron Microscopy},
  journal = {Science},
  volume  = {328},
  pages   = {187--193},
  year    = {2010},
  doi     = {10.1126/science.1166135}
}

@article{Sciaini2011RPP,
  author  = {Sciaini, Germ{\'a}n and Miller, R. J. Dwayne},
  title   = {Femtosecond electron diffraction: heralding the era of atomically resolved dynamics},
  journal = {Reports on Progress in Physics},
  volume  = {74},
  pages   = {096101},
  year    = {2011},
  doi     = {10.1088/0034-4885/74/9/096101}
}

@article{Filippetto2022RMP,
  author  = {Filippetto, Daniele and Musumeci, Pietro and Li, Renkai and Siwick, Bradley J. and Otto, Martin R. and Centurion, Martin and Nunes, Joao Pedro F.},
  title   = {Ultrafast electron diffraction: Visualizing dynamic states of matter},
  journal = {Reviews of Modern Physics},
  volume  = {94},
  pages   = {045004},
  year    = {2022},
  doi     = {10.1103/RevModPhys.94.045004}
}

@article{Morimoto2017JAP,
  author  = {Morimoto, Yuya and Roland, I. and Rennesson, S. and Semond, F. and Boucaud, P. and Baum, Peter},
  title   = {Laser damage of free-standing nanometer membranes},
  journal = {Journal of Applied Physics},
  volume  = {122},
  pages   = {215303},
  year    = {2017},
  doi     = {10.1063/1.5006906}
}

@article{Kirchner2014NatPhotonics,
  author  = {Kirchner, F. O. and Gliserin, A. and Krausz, F. and Baum, P.},
  title   = {Laser streaking of free electrons at 25 keV},
  journal = {Nature Photonics},
  volume  = {8},
  pages   = {52--57},
  year    = {2014},
  doi     = {10.1038/nphoton.2013.315}
}

@article{Priebe2017NatPhotonics,
  author  = {Priebe, K. E. and Rathje, C. and Yalunin, S. V. and Hohage, T. and Feist, A. and Sch{\"a}fer, S. and Ropers, C.},
  title   = {Attosecond electron pulse trains and quantum state reconstruction in ultrafast transmission electron microscopy},
  journal = {Nature Photonics},
  volume  = {11},
  pages   = {793--797},
  year    = {2017},
  doi     = {10.1038/s41566-017-0045-8}
}

@article{Ehberger2018PRL,
  author  = {Ehberger, D. and Ryabov, A. and Baum, P.},
  title   = {Tilted Electron Pulses},
  journal = {Physical Review Letters},
  volume  = {121},
  pages   = {094801},
  year    = {2018},
  doi     = {10.1103/PhysRevLett.121.094801}
}

@article{Ehberger2019PRApplied,
  author  = {Ehberger, D. and Mohler, K. J. and Vasileiadis, T. and Ernstorfer, R. and Waldecker, L. and Baum, P.},
  title   = {Terahertz compression of electron pulses at a planar mirror membrane},
  journal = {Physical Review Applied},
  volume  = {11},
  pages   = {024034},
  year    = {2019},
  doi     = {10.1103/PhysRevApplied.11.024034}
}

@article{Ryabov2020SciAdv,
  author  = {Ryabov, A. and Thurner, J. W. and Nabben, D. and Tsarev, M. V. and Baum, P.},
  title   = {Attosecond metrology in a continuous-beam transmission electron microscope},
  journal = {Science Advances},
  volume  = {6},
  pages   = {eabb1393},
  year    = {2020},
  doi     = {10.1126/sciadv.abb1393}
}

@article{Vanacore2018NatCommun,
  author  = {Vanacore, G. M. and Madan, I. and Berruto, G. and Wang, K. and Pomarico, E. and Lamb, R. J. and McGrouther, D. and Kaminer, I. and Barwick, B. and Garc{\'i}a de Abajo, F. J. and Carbone, F.},
  title   = {Attosecond coherent control of free-electron wave functions using semi-infinite light fields},
  journal = {Nature Communications},
  volume  = {9},
  pages   = {2694},
  year    = {2018},
  doi     = {10.1038/s41467-018-05021-x}
}

@article{Tsarev2021PRResearch,
  author  = {Tsarev, Maxim V. and Ryabov, Andrey and Baum, Peter},
  title   = {Free-Electron Qubits and Maximum-Contrast Attosecond Pulses via Temporal {Talbot} Revivals},
  journal = {Physical Review Research},
  volume  = {3},
  pages   = {043033},
  year    = {2021},
  doi     = {10.1103/PhysRevResearch.3.043033}
}

@article{Feist2020PRResearch,
  author  = {Feist, A. and Yalunin, S. V. and Sch{\"a}fer, S. and Ropers, C.},
  title   = {High-purity free-electron momentum states prepared by three-dimensional optical phase modulation},
  journal = {Physical Review Research},
  volume  = {2},
  pages   = {043227},
  year    = {2020},
  doi     = {10.1103/PhysRevResearch.2.043227}
}

@article{RoadmapFreeElectrons2025ACSPhotonics,
  author  = {Garc{\'i}a de Abajo, F. Javier and Polman, Albert and Velasco, Cruz I. and Kociak, Mathieu and Tizei, Luiz H. G. and St{\'e}phan, Odile and Meuret, Sophie and Sannomiya, Takumi and Akiba, Keiichiro and Auad, Yves and others},
  title   = {Roadmap for Quantum Nanophotonics with Free Electrons},
  journal = {ACS Photonics},
  volume  = {12},
  number  = {9},
  pages   = {4760--4817},
  year    = {2025},
  doi     = {10.1021/acsphotonics.5c00585}
}

@article{Ferrari2025ACSPhotonics,
  author  = {Ferrari, Beatrice Matilde and Duncan, Cameron James Richard and Ostroman, Irene and Bravi, Maria Giulia and Rosi, Paolo and others},
  title   = {Realization of a Pre-Sample Photonic-Based Free-Electron Modulator in Ultrafast Transmission Electron Microscopes},
  journal = {ACS Photonics},
  volume  = {12},
  number  = {11},
  pages   = {5864--5873},
  year    = {2025},
  doi     = {10.1021/acsphotonics.5c00549}
}

@article{Morimoto2022AnnPhys,
  author  = {Morimoto, Yuya and Chen, Bo-Han and Baum, Peter},
  title   = {Free-electron tomography of few-cycle optical waveforms},
  journal = {Annalen der Physik},
  volume  = {534},
  pages   = {2200193},
  year    = {2022},
  doi     = {10.1002/andp.202200193}
}

@article{Nabben2023Nature,
  author  = {Nabben, D. and Kuttruff, J. and Stolz, L. and Ryabov, A. and Baum, P.},
  title   = {Attosecond electron microscopy of sub-cycle optical dynamics},
  journal = {Nature},
  volume  = {619},
  pages   = {63--67},
  year    = {2023}
}

@article{Gaida2024NatPhotonics,
  author  = {Gaida, J. H. and Louren{\c{c}}o-Martins, H. and Sivis, M. and Rittmann, T. and Feist, A. and Garc{\'i}a de Abajo, F. J. and Ropers, C.},
  title   = {Attosecond electron microscopy by free-electron homodyne detection},
  journal = {Nature Photonics},
  volume  = {18},
  pages   = {509--515},
  year    = {2024}
}

@article{Bucher2024NatPhotonics,
  author  = {Bucher, T. and Nahari, H. and Herzig Sheinfux, H. and Ruimy, R. and Niedermayr, A. and Dahan, R. and Yan, Q. and Adiv, Y. and Yannai, M. and Chen, J. and others},
  title   = {Coherently amplified ultrafast imaging using a free-electron interferometer},
  journal = {Nature Photonics},
  volume  = {18},
  pages   = {809--817},
  year    = {2024}
}

@article{Morimoto2024PRL,
  author  = {Morimoto, Yuya and Baum, Peter},
  title   = {Field-induced rocking-curve effects in attosecond electron diffraction},
  journal = {Physical Review Letters},
  volume  = {132},
  pages   = {216902},
  year    = {2024},
  doi     = {10.1103/PhysRevLett.132.216902}
}

@article{Kroll1976PRA,
  author  = {Kroll, Norman M. and Watson, Kenneth M.},
  title   = {Inelastic Atom-Atom Scattering within an Intense Laser Beam},
  journal = {Physical Review A},
  volume  = {13},
  pages   = {1018--1025},
  year    = {1976},
  doi     = {10.1103/PhysRevA.13.1018}
}

@article{Weingartshofer1977PRL,
  author  = {Weingartshofer, A. and Holmes, J. K. and Caudle, G. and Clarke, E. M. and Kruger, H.},
  title   = {Direct Observation of Multiphoton Processes in Laser-Induced Free-Free Transitions},
  journal = {Physical Review Letters},
  volume  = {39},
  pages   = {269--272},
  year    = {1977},
  doi     = {10.1103/PhysRevLett.39.269}
}

@article{Freimund2001Nature,
  author  = {Freimund, Daniel L. and Aflatooni, Kayvan and Batelaan, Herman},
  title   = {Observation of the {Kapitza--Dirac} Effect},
  journal = {Nature},
  volume  = {413},
  pages   = {142--143},
  year    = {2001},
  doi     = {10.1038/35093065}
}

@article{Plettner2005PRL,
  author  = {Plettner, Tomas and Byer, Robert L. and Colby, Eric and Cowan, Benjamin and Sears, Christopher M. S. and Spencer, James E. and Siemann, Robert H.},
  title   = {Visible-Laser Acceleration of Relativistic Electrons in a Semi-Infinite Vacuum},
  journal = {Physical Review Letters},
  volume  = {95},
  pages   = {134801},
  year    = {2005},
  doi     = {10.1103/PhysRevLett.95.134801}
}

@article{Madan2022ACSPhotonics,
  author  = {Madan, Ivan and Leccese, Veronica and Mazur, Adam and Barantani, Francesco and LaGrange, Thomas and Sapozhnik, Alexey and Tengdin, Phoebe M. and Gargiulo, Simone and Rotunno, Enzo and Olaya, Jean-Christophe and Kaminer, Ido and Grillo, Vincenzo and Garc{\'i}a de Abajo, F. Javier and Carbone, Fabrizio and Vanacore, Giovanni Maria},
  title   = {Ultrafast Transverse Modulation of Free Electrons by Interaction with Shaped Optical Fields},
  journal = {ACS Photonics},
  volume  = {9},
  number  = {10},
  pages   = {3215--3224},
  year    = {2022},
  doi     = {10.1021/acsphotonics.2c00850}
}

@article{Meuret2024ACSPhotonics,
  author  = {Meuret, Sophie and Louren{\c{c}}o-Martins, Hugo and Weber, S{\'e}bastien J. and Houdellier, Florent and Arbouet, Arnaud},
  title   = {Photon-Induced Near-Field Electron Microscopy of Nanostructured Metallic Films and Membranes},
  journal = {ACS Photonics},
  volume  = {11},
  number  = {3},
  pages   = {977--984},
  year    = {2024},
  doi     = {10.1021/acsphotonics.3c01223}
}

@article{Wang2024PRB,
  author  = {Wang, Wentao and Zheng, Dingguo and Huang, Siyuan and Li, Jun and others},
  title   = {Energy-Momentum Transfer in the Free-Electron--Photon Interaction Mediated by a Film},
  journal = {Physical Review B},
  volume  = {109},
  pages   = {134305},
  year    = {2024},
  doi     = {10.1103/PhysRevB.109.134305}
}

@article{Muller2024Arxiv,
  author  = {M{\"u}ller, N. and el Kabil, S. and Vosse, G. and Hansen, L. and Rathje, C. and Sch{\"a}fer, S.},
  title   = {Spectrally resolved free electron-light coupling strength in a transition metal dichalcogenide},
  journal = {arXiv preprint},
  year    = {2024},
  eprint  = {2405.12017},
  archivePrefix = {arXiv},
  primaryClass  = {cond-mat.mes-hall}
}

@article{Morimoto2018PRA,
  author  = {Morimoto, Yuya and Baum, Peter},
  title   = {Attosecond control of electron beams at dielectric and absorbing membranes},
  journal = {Physical Review A},
  volume  = {97},
  pages   = {033815},
  year    = {2018},
  doi     = {10.1103/PhysRevA.97.033815}
}

@book{Hecht2016Optics,
  author    = {Hecht, Eugene},
  title     = {Optics, Global Edition},
  edition   = {5},
  publisher = {Pearson Education Limited},
  year      = {2016},
  isbn      = {9781292096933}
}

@article{MorimotoBaum2018NatPhys,
  author  = {Morimoto, Yuya and Baum, Peter},
  title   = {Diffraction and microscopy with attosecond electron pulse trains},
  journal = {Nature Physics},
  volume  = {14},
  pages   = {252--256},
  year    = {2018},
  doi     = {10.1038/s41567-017-0007-6}
}

@article{Kozak2018PRL,
  author  = {Koz{\'a}k, M. and Sch{\"o}nenberger, N. and Hommelhoff, P.},
  title   = {Ponderomotive generation and detection of attosecond free-electron pulse trains},
  journal = {Physical Review Letters},
  volume  = {120},
  pages   = {103203},
  year    = {2018},
  doi     = {10.1103/PhysRevLett.120.103203}
}

@article{Morimoto2015PRL,
  author  = {Morimoto, Yuya and Kanya, Reika and Yamanouchi, Kaoru},
  title   = {Light-dressing effect in laser-assisted elastic electron scattering by {Xe}},
  journal = {Physical Review Letters},
  volume  = {115},
  pages   = {123201},
  year    = {2015},
  doi     = {10.1103/PhysRevLett.115.123201}
}

@article{Tachibana2026,
  author  = {Yuichi Tachibana and Marie Ouill{\'e} and Takuya Hosobata
             and Takashi Isoshima and Yoshiyuki Takizawa and Yutaka Yamagata
             and Yuya Morimoto},
  title   = {Attosecond shaping of multi-electron pulses in a home-built
             37-keV beamline},
  journal = {Scientific Reports},
  year    = {2026},
  doi     = {10.1038/s41598-026-63961-7},
  url     = {https://doi.org/10.1038/s41598-026-63961-7}
}

@book{Joachain2012Book,
  author    = {C. J. Joachain and N. J. Kylstra and R. M. Potvliege},
  title     = {Atoms in Intense Laser Fields},
  publisher = {Cambridge University Press},
  address   = {Cambridge, UK},
  year      = {2012},
  doi       = {10.1017/CBO9780511993459},
  isbn      = {9780521793018}
}

@article{Morimoto2014JCP,
  author  = {Yuya Morimoto and Reika Kanya and Kaoru Yamanouchi},
  title   = {Laser-assisted electron diffraction for femtosecond molecular
             imaging},
  journal = {J. Chem. Phys.},
  volume  = {140},
  number  = {6},
  pages   = {064201},
  year    = {2014},
  doi     = {10.1063/1.4863985}
}

@article{Taleb2025NatCommun,
  author  = {M. Taleb and P. H. Bittorf and M. Black and M. Hentschel
             and W. Sigle and B. Haas and C. Koch and P. A. van Aken
             and H. Giessen and N. Talebi},
  title   = {Ultrafast phonon-mediated dephasing of color centers in
             hexagonal boron nitride probed by electron beams},
  journal = {Nat. Commun.},
  volume  = {16},
  pages   = {2326},
  year    = {2025},
  doi     = {10.1038/s41467-025-57584-1}
}
\end{document}